# Observing Li Nucleation at Li Metal–Solid Electrolyte Interface in All-Solid-State Batteries

*Yun An,[1] Taiping Hu,[1,2] Quanquan Pang,[1*] and Shenzhen Xu[1,2*]*

[1]Beijing Key Laboratory for Theory and Technology of Advanced Battery Materials, School of Materials Science and Engineering, Peking University, Beijing 100871, China

[2]AI for Science Institute, Beijing 100084, China

E-mail: *qqpang@pku.edu.cn

*xushenzhen@pku.edu.cn

Benefiting from the significantly improved energy density and safety, all-solid-state lithium batteries (ASSLBs) are considered one of the most promising next-generation energy technologies. Their practical applications, however, are strongly impeded by the Li dendrite formation. Despite this recognized challenge, a comprehensive understanding of Li dendrite nucleation and formation mechanism remains elusive. In particular, the initial locations of Li dendrite formation are still ambiguous: do Li clusters form directly at the Li anode surface, or inside the bulk solid electrolyte (SE), or within the solid-electrolyte interphase (SEI)? Here, based on the deep-potential molecular dynamics simulations combined with enhanced sampling techniques, we investigate the atomic-level mechanism of Li cluster nucleation and formation at the Li anode/SE interface. We observe that an isolated Li cluster initially forms inside the SEI between the $Li_6PS_5Cl$ SE and the Li metal anode, located ~1 nm away from the Li anode/SEI boundary. The local electronic structure of the spontaneously formed SEI is found to be a key factor enabling the Li cluster formation within SEI, in which a significantly decreased bandgap could facilitate electronic conduction through the SEI and reduce $Li^+$ ions to metallic Li atoms therein. Our work therefore provides atomic-level insights into Li-dendrite nucleation at anode/SE interfaces in ASSLBs, and could guide future design for developing Li-dendrite-inhibiting strategies.

## 1. Introduction

All-solid-state lithium batteries (ASSLBs), using lithium (Li) metal as anode and replacing liquid electrolytes with non-flammable solid electrolytes (SEs), are considered one of the most promising next-generation energy technologies due to their significantly improved energy density and safety compared to conventional Li-ion batteries.[1-3] Nevertheless, the successful commercialization of ASSLBs faces tremendous challenges,[4,5] prominently by the inevitable formation of Li dendrites that lead to substantial safety hazards, diminished energy density, and limited operational lifespans.[6,7]

Despite the extensive efforts dedicated to understanding the origin of lithium dendrites within ASSLBs,[8-11] a fully recognized mechanism of Li dendrite nucleation and formation remains elusive. In general, Li dendrites were found to form at the Li anode/SE interface, i.e., initially directly connected to the Li anode,[12,13] and as they further grow, eventually penetrate the SE and cause short circuits. Voids, contact loss, and nano-cracks were considered to be the dominant reasons for these anode-initiated dendrites.[13,14] Some recent experimental reports, however, observed that Li dendrites form inside SEs,[15-18] i.e., the dendrites do not necessarily have a direct connection to the Li anode. The electronic conductivity of SE and grain boundaries (GBs), were demonstrated to be related to the internal Li dendrite formation. For instance, Li dendrites were observed inside the bulk $Li_7La_3Zr_2O_{12}$ (LLZO) and $Li_3PS_4$ due to the high electronic conductivity of these SE materials;[15] GBs could provide possible electronic leakage channels and result in $Li^+$ ions reduction inside the SE.[16,19] Although considerable progress has been made, the mechanism of Li dendrite formation is still under debate, particularly regarding the initial locations of Li dendrite formation. To identify the initial growth sites of Li dendrites and enable designing dendrite-free batteries, several fundamental questions must be addressed: (1) Are the initially formed

Li clusters directly connected to the Li anode, or located at a distance from the Li anode/SE interface within the solid-electrolyte interphase (SEI) region, or even inside the SE bulk? (2) If not directly connected to the Li anode, what is the channel of electronic transport to allow reduction of $Li^+$ ions to metallic Li atoms at the distance?

To answer these questions, in this paper, we investigate the atomic-level mechanism of Li dendrite/cluster nucleation and formation in a Li anode/SE interfacial system by employing molecular dynamics (MD) simulations. The argyrodite $Li_6PS_5Cl$,[20] a promising superionic conductor, is used as the SE material in this work. Given that capturing even the initial Li dendrite/cluster formation process requires large time scale simulations, which are beyond the ability of expensive first-principles calculations, we employ machine learning techniques and train a deep potential (DP) [21] with an accuracy benchmarked with first-principles calculations. In addition, in view that conventional MD simulations face challenges in capturing dynamics of high-barrier Li cluster formation processes within an affordable computational time, we employ an enhanced-sampling technique in the deep potential molecular dynamics (DPMD) simulations. Our enhanced-sampling MD trajectories show that an isolated Li cluster in fact forms inside the SEI region, approximately 1 nm away from the Li anode/SEI interfacial boundary, rather than directly grows connected to the Li anode surface or within the SE bulk. A high electronic conductivity of the spontaneously formed SEI layer, revealed by our first-principles electronic structure calculations, reconciles the necessity of electronic conduction for Li reduction. Understanding the initial positions of Li dendrite/cluster formation and the electronic conductivity of the SEI layer between the anode and the SE are crucial for designing strategies of suppressing dendrite generation.

## 2. Methods

### 2.1. Machine learning-based potential training

We employed an automatic workflow, the Deep Potential GENerator (DP-GEN)[22]

scheme (**Figure 1**), to train a deep potential. Prior to the DP-GEN iterations, we generated an initial dataset by using density functional theory (DFT) calculations with randomly perturbated structures, including a bulk body-centered cubic (BCC) Li phase (16 atoms), a bulk $Li_6PS_5Cl$ phase (52 atoms), and an interfacial model of $Li/Li_6PS_5Cl$ (185 atoms) (Figure 1a). Less atoms in these small-size systems enable efficient DFT calculations, which were performed using the *Vienna Ab initio Simulation Package* (VASP).[23,24] The core-valence electron interaction was described by the projector augmented wave (PAW) approach,[25,26] and we employed the Perdew-Burke-Ernzerhof[27] (PBE) functional to represent the electronic exchange-correlation effect. We set the kinetic cutoff energy to be 600 eV for the plane-wave basis. The convergence criteria of energies and forces were set to be $10^{-6}$ eV and 0.01 eV/Å, respectively. Please refer to Supporting Information (SI) for more computational details. We then used the dataset produced by DFT for labeling in the DP-GEN iterations to train the DP potential (Figure 1b). Subsequently, the trained DP was applied in the enhanced-sampling DPMD simulations with a symmetric cell of $Li|Li_6PS_5Cl|Li$, consisting of 4833 atoms in total (Figure 1c). The bottom region in the model of $Li|Li_6PS_5Cl|Li$ is a BCC-Li slab as an anode (16 Li layers); the middle region is the $Li_6PS_5Cl$ electrolyte; and the upper material of BCC Li (9 Li layers) is included in our interfacial supercell to represent a source of Li particles from an assumed deeper SE region. We also insert a vacuum region between the anode Li slab and the Li source slab to mimic a half-cell interfacial system (illustrated in Figure 1c). More details about the atomic model constructions are given in the SI.

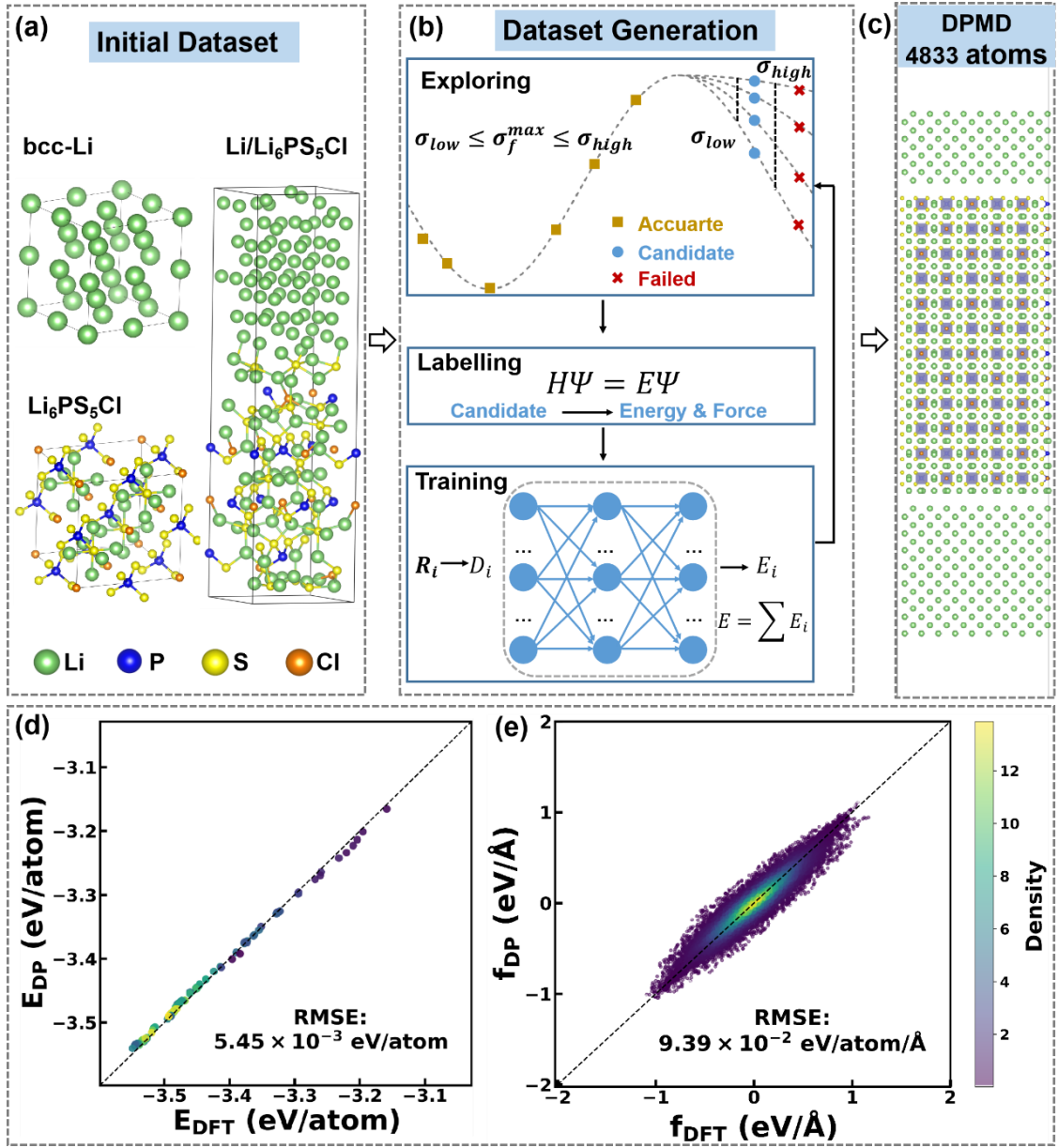

**Figure 1.** DP-GEN workflow for training the DP potential used in this study. (a) Initial dataset construction. (b) DP-GEN iterations consisting of configurational exploration, DFT labeling, and neural network training. (c) The Li|Li$_6$PS$_5$Cl|Li supercell (containing 4833 atoms) constructed for enhanced-sampling DPMD simulations. DPtest results of (d) energies and (e) forces predicted from DP inferences and DFT calculations on the testing dataset of the Li/Li$_6$PS$_5$Cl interfacial model.

### 2.2. Collective variable setup in enhanced-sampling MD simulations

Li dendrite/cluster formation typically needs to overcome high kinetic barriers, and conventional unbiased MD simulations usually fail to capture this process within an acceptable computational time. Therefore, we employed an enhanced sampling technique[28] - the moving restraint method,[29] in the DPMD simulations (see SI for more details). This method adds a time-dependent restraint, via a harmonic potential of $\frac{1}{2}\kappa(s(\vec{R}) - s_0(t))^2$, on defined Collective Variables (*CV*s). Here $\kappa$ is the force constant, $s$ is the defined *CV* as a function of the system's atomic configuration $\vec{R}$, and the time-dependent $s_0(t)$ performs as a knob to move the *CV* along a specific direction at a specific rate (see SI for details of these parameters' setup). As such, it allows an efficient configurational sampling towards a certain target in terms of the *CV*'s evolvement, which connects an initial state to our interested final state,[30] corresponding to Li dendrite/cluster nucleation process in our work.

We employed atoms' coordination numbers (*CN*) to construct the *CV* in our enhanced-sampling DPMD simulations. The coordination numbers of Li-Li, Li-S, Li-P, and Li-Cl were used in the modeled system. The *CN* quantity[31] follows the switching function of

$$CN_{AB} = \sum_{i \in A} \sum_{j \in B} \frac{\left(1 - \frac{r_{ij}}{R_0}\right)^n}{\left(1 - \frac{r_{ij}}{R_0}\right)^m} \tag{1}$$

where the $r_{ij}$ is the distance between the $i^{th}$ atom among the species A and $j^{th}$ atom among the species B; the exponents *n* and *m* determine the sharpness of the switching function; $R_0$ represents a spherical radius for defining a neighboring region of a centered atom (belonging to species A) when counting coordinated atoms (within species B).[31] For $CN_{\text{Li-Li}}$, species A included all the Li atoms in $Li_6PS_5Cl$ SE, while species B included the upper 8 layers of Li atoms from anode Li slab in addition to all the Li atoms in $Li_6PS_5Cl$ SE (**Figure 2**a). For $CN_{\text{Li-S}}$, $CN_{\text{Li-P}}$, and $CN_{\text{Li-Cl}}$, species A consisted of all the Li atoms in $Li_6PS_5Cl$ SE, while species B consisted of S, P, and Cl atoms in the $Li_6PS_5Cl$ SE, respectively. The detailed setup of *CN* parameters is provided in Table S1 of the SI.

We used a linear combination of the *CN* to construct the *CV* (Equation (2)), which would be driven by the moving restraint algorithm in our enhanced-sampling MD simulations, in which the $s_0(t)$ (defined in the harmonic restraint shown in the above paragraph) gradually changes from an initial-state value to a final-state value, enabling us to observe the system's structural evolution along a process with a high kinetic barrier. Regarding the reasons for choosing the above *CV* defined in Equation (2) below, we consider the following underlying principles:

$$CV = CN_{\text{Li}-\text{Li}} - CN_{\text{Li}-\text{S}} - CN_{\text{Li}-\text{P}} - CN_{\text{Li}-\text{Cl}} \qquad (2)$$

(1) The primary feature of Li dendrite/cluster nucleation is the change of the Li coordination number, i.e. variation in its coordination environment with different surrounding elements (e.g., Li-Li/S/P/Cl). Therefore, we used the coordination number of Li-Li/S/P/Cl in the *CV*, as the key indicator of modeling Li clusters formation and growth.

(2) During the initial stage of Li dendrite/cluster nucleation, some Li atoms, originally coordinated with S, Cl, P anions, would gradually transform to local environment with Li-Li coordination. We thus need a *CV* that reflects an increase in Li-Li coordination number while decrease in coordination with Li-S/P/Cl anions when dragging the *CV* value towards a certain direction.

(3) As we do not know whether the Li dendrite/cluster would initially form with a direct connection with the Li anode surface, or inside the SE bulk, or within the SEI layer (isolated from the Li anode surface with a finite distance), we need to include the Li atoms that can represent these three scenarios in *CV* when we count the Li-Li coordination numbers. That is, we use Li atoms belonging to the anode slab and Li$_6$PS$_5$Cl SE bulk to count the Li-Li coordination numbers, thus enabling the possibility that if the dendrite growth right from the Li anode surface is energetically competitive, Li clustering with a direct connection with the Li metal surface would contribute to catch up with the gradual *CV* variation along the moving-restraint MD trajectory. Therefore, we include the top 8 layers of Li atoms from the Li anode slab in species B defined in Equation (1), to determine whether Li cluster initially grows directly along the Li anode surface or at a distance within the SEI. In addition, including Li atoms from the Li$_6$PS$_5$Cl SE allows us to examine whether Li cluster forms within the SE bulk. Overall, our *CV* effectively covers all three possible Li cluster formation mechanisms.

With these key factors (discussed above) being taken account of, if we gradually increase the *CV* value, we can imagine that the system's configuration would be driven to a state with more Li-Li coordinated local environment while decreasing the Li-S/Cl/P bonds, whether along a process with Li clustering right from the Li anode surface, or within the bulk of the SE, or through a path with Li clustering within the SEI inner region (not directly connected to the Li metal surface). The MD simulation results thus can tell us which mechanism is more likely to occur at the Li anode/SE interface, which is the key scientific question investigated in this work. It is important to note that, however, the electrochemical potential of electrodes is the major driving force for Li dendrite/cluster formation in realistic operation conditions, via a redox process Li$^+$ + e$^-$ → Li$^0$. In this work, we acknowledge that this driving force is assumed to be represented by the *CV* variation along the enhanced-sampling MD trajectories, to achieve practical simulations of Li dendrite/cluster nucleation dynamics.

**2.3. Enhanced-sampling MD simulations**

We performed enhanced-sampling MD simulations via the interface of the deepmd-kit[22] within the LAMMPS package,[32] using a Li|Li$_6$PS$_5$Cl|Li supercell model containing 4833 atoms (initial box lengths: 30.81×30.81×150.42 Å$^3$, Figure 1c). We employed the NPT ensemble condition in our simulations, allowing the volume to relax in response to an ambient pressure. We set the pressures to be 1 bar along the *x* and *y* directions while the cell's length along the *z* direction is fixed due to the presence of vacuum region (no pressure was set along this *z* direction). The periodic boundary condition was applied for all of the three directions. We used the Nosé-Hoover thermostat coupled with the Parrinello-Rahman dynamics[33] to achieve the NPT condition. The temperature was controlled as 300 K. We set the time step to be 1 fs to generate particle trajectories via MD time integration. The total simulation time was

150 ps. Given that we employed enhanced sampling techniques in the MD simulations, the simulation time here does not have actual physical meaning of dynamics and was only used to represent the total number of MD steps.

**2.4. Charge equilibration method**

We employed the charge equilibration (QEq) approach[34,35] to estimate the equilibrium charge distributions in our system, enabling qualitative evaluations of Li atomic charge variations during the enhanced-sampling MD simulations. Since the Li valence state in Li clusters or dendrites is similar to that in a metallic phase (Li$^0$), in a sharp contrast to the Li valence state in SE or SEI phases (Li$^+$), we thus can utilize the QEq method to perform Li atomic charge analysis, and to identify the formation of Li cluster or dendrite in the complicated interfacial system. In the QEq method, for a system containing $N$ atoms, the atomic charges $Q_i$ are optimized to minimize the QEq energy, $E_{\text{QEq}}$[35]:

$$E_{\text{QEq}} = E_{\text{Coulomb}} + \sum_i^N \left( \chi_i^0 Q_i + \frac{1}{2} J_i Q_i^2 \right) \qquad (3)$$

where $E_{\text{Coulomb}}$ is the Coulomb interactions, which we calculated through the Particle Mesh Ewald (PME) [36] algorithm (Equation (S3) in SI), along with additional correction for the Gaussian charge distribution (Equation (S7) in SI). $\chi_i^0$, $Q_i$, and $J_i$ are the electronegativity, atomic charge, and atomic hardness, respectively. The parameters for $\chi_i^0$ and $J_i$ for each element were set at the beginning of molecular simulations (Table S2 in the SI).

The sum of all atomic charges in the system is constrained to the total charge $Q_{\text{tot}}$, i.e., $\sum_{i=1}^N Q_i = Q_{\text{tot}}$. To solve this constrained minimization problem, we used the Lagrange multiplier method:

$$\mathcal{L} = E_{\text{QEq}} - \chi_{\text{eq}} \left( \sum_{i=1}^N Q_i - Q_{\text{tot}} \right) \qquad (4)$$

The QEq charges $\{Q_i\}$ and the Lagrange multiplier $\chi_{\text{eq}}$ can then be determined by

$$\begin{cases} \dfrac{\partial \mathcal{L}}{\partial Q_i} = 0 \\ \dfrac{\partial \mathcal{L}}{\partial \chi_{\text{eq}}} = 0 \end{cases} \quad (5)$$

The atomic charges $Q_i$ in the subsequent calculations were obtained based on the above method. For more details on the QEq, please refer to the SI.

## 3. Results and Discussion

We first validate our trained DP model by comparing energies and forces predicted by the DP force field with those calculated by the DFT approach. This validation is performed on a Li-Li$_6$PS$_5$Cl interfacial system containing 185 atoms (right panel of Figure 1a). Using our trained DP model, we conduct 200 ps of NPT DPMD simulations and extract trajectories to construct the testing dataset with 100 configurations randomly selected from the trajectories. We then use the DFT simulation package VASP[23,24] to calculate the single-point energies of these configurations (see SI for more details of DFT). Energy and force comparisons on the testing dataset exhibit a good agreement between our trained DP model and the DFT results, with a root-mean-square error (RMSE) of 5.45 meV/atom for energies and 93.9 meV/Å for forces, respectively (Figure 1d and e), justifying the accuracy of our DP model. For the tests on the training dataset, please refer to Figure S1 in SI.

We then perform enhanced-sampling DPMD simulations for the Li|Li$_6$PS$_5$Cl|Li interfacial supercell at $T = 300$ K. During the initial stage of MD simulations (~ 0 – 100 ps), we find the upper slab of BCC Li (above the Li$_6$PS$_5$Cl) turns amorphous and some Li atoms transport downward into the SE, which agrees with our assumption of serving as the Li source. For the bottom BCC Li anode, the major part of the slab maintains its bulk feature, except for a few layers close to the interface that interact with the Li$_6$PS$_5$Cl, forming the SEI region between Li anode and Li$_6$PS$_5$Cl. We can clearly observe that Li-S, Li-P, and Li-Cl bonds are largely present in the SEI (Figure 2b), corresponding to the decomposition components such as Li$_2$S, Li$_3$P, and LiCl, which are likely to be

produced by the reaction of $Li_6PS_5Cl + 8Li \rightarrow 5Li_2S + Li_3P + LiCl$, as proposed by previous work.[37,38] We can also see that $Li_6PS_5Cl$ maintains its bulk crystalline phase in the central region of SE region after $10^5$ MD steps (100 ps).

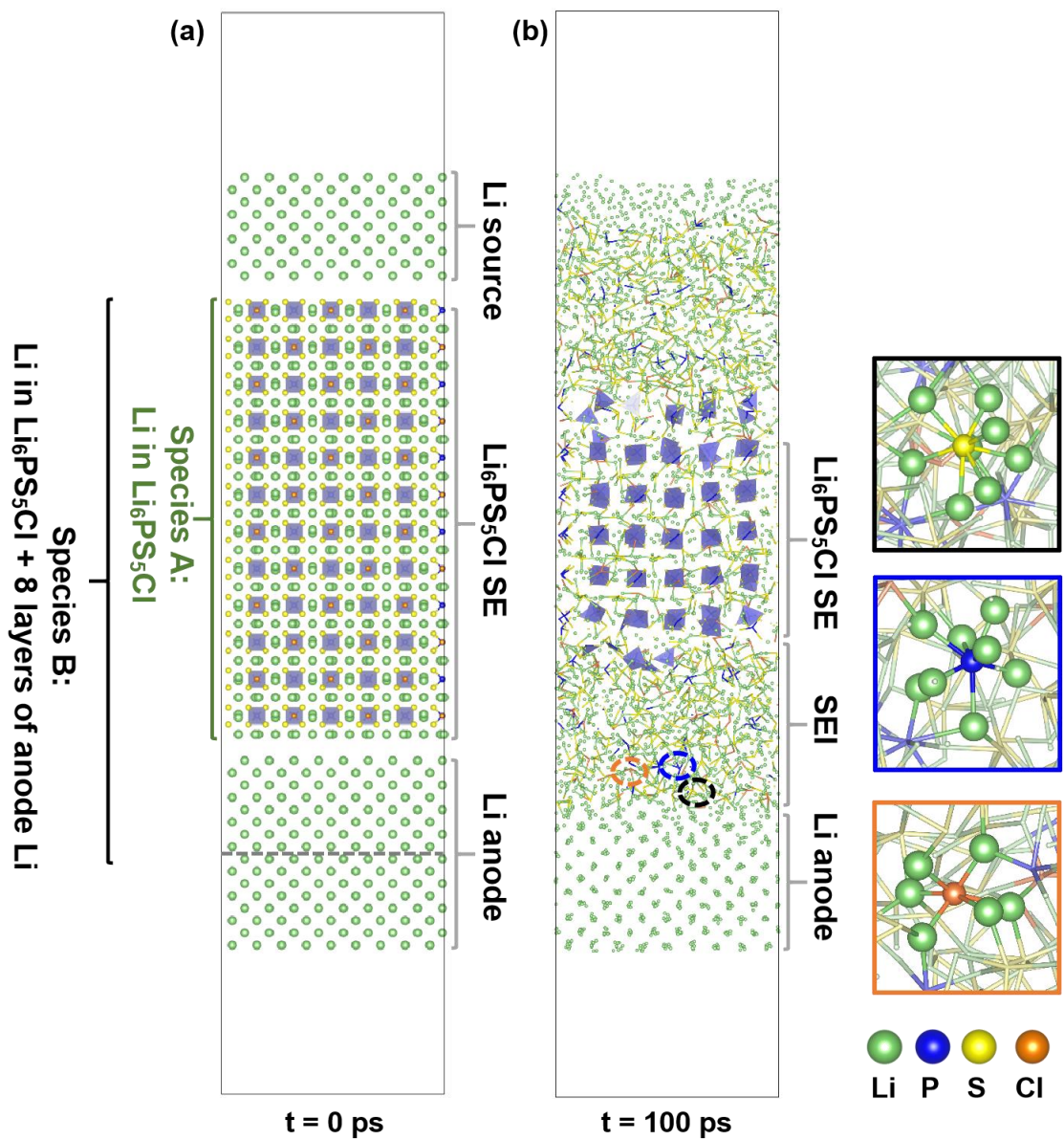

**Figure 2.** (a) Evolution of the Li|Li$_6$PS$_5$Cl|Li system along enhanced-sampling DPMD simulations at 0 ps; the green and black brackets indicate the regions of species A and B used to set the Li-Li coordination numbers in the *CV*. Green, blue, yellow, and orange spheres denote Li, P, S, and Cl atoms, respectively. (b) Evolution of the Li|Li$_6$PS$_5$Cl|Li system along enhanced-sampling DPMD simulations at 100 ps. The black, blue, and orange dashed circles highlight the newly formed Li-S, Li-P, and Li-Cl bonds, which are further illustrated by the zoom-in structural plots on the right side.

As the simulations proceed, the thickness of the SEI increases, indicating the consumption of both the SE and the Li anode. Of note, we observe an emergence of a Li cluster within the SEI region (atoms highlighted in wine color in **Figure 3**a) at ~125 ps, indicating nucleation of a metallic Li phase. This cluster continues to grow, becoming more distinct by 150 ps. To identify the process of Li clusters nucleation, we calculate the QEq charges for the configurations at 100, 125 and 150 ps and analyze the Li atomic charge distribution, with the color-coded charge values shown in Figure 3b-d. The Li atoms in the Li anode region, with charges close to 0, are represented in blue, while $Li^+$ ions in the bulk phase of $Li_6PS_5Cl$, with charges close to +1, are shown in red. Most Li within the SEI exhibit intermediate charges between 0 and +1, are represented in green. At 100 ps when no Li cluster is present in the system, the Li atoms in the SEI exhibit intermediate charges, shown in green in Figure 3b. In contrast, at 125 and 150 ps, as the Li cluster starts to nucleate and grow within the SEI, a small group of Li atoms in SEI appear blue (Figure 3c,d), confirming the formation of a Li cluster within the SEI inner region. In addition, instead of being directly connected to the Li anode, the emerged Li cluster is located ~ 1 nm away from the Li/SEI interface, consistent with previous experimental findings.[17] It is important to note that the Li cluster neither initiate within the bulk of the $Li_6PS_5Cl$ SE. To understand the origin of the Li cluster nucleation inside the SEI region, we track the initial positions of these clustered Li atoms and find that the Li atoms are mainly from the SEI region (produced by the side reactions of Li metal with the SE during our MD simulations). If we further track back to the initial state of our modeled Li|$Li_6PS_5Cl$|Li interfacial system, the Li atoms in the nucleated Li cluster (at 125 ps) come from the $Li_6PS_5Cl$ SE at the beginning of our MD trajectories. Radial distribution function (RDF) analysis of the clustered Li atoms (marked by the red circle in **Figure 4**a) shows that the distance between the nearest Li-Li neighbor is 2.75 Å (red line in Figure 4b), slightly shorter than that in BCC-Li (~2.96 Å, blue line in Figure 4b).

The above results suggest that the initial growth of Li dendrites is more likely to occur through Li clustering within the inner region of the SEI, rather than directly from the

Li metal surface at the Li anode/SE interface or within the SE bulk. In principle, our designed *CV* also allows Li dendrite growth right along the anode/SEI surface and within the SE bulk, both of which could effectively increase the *CV* as well when we drag the *CV* along the moving restraint MD path. In particular, in the configurations from our enhanced-sampling MD simulations where Li cluster emerges, either 125 ps or 150 ps, the upper 3 layers of Li anode slab (above the gray dashed line in Figure 3a) remain within the region of species B in the *CN* defined in Equation (1). Therefore, these Li atoms in the anode still can contribute when counting the Li-Li coordination numbers, which would potentially trigger Li dendrite formation at the Li anode/surface. However, while we well provide such possibilities, our MD simulations results show no Li dendrite formation directly connected to the Li anode. Instead, Li clustering emerges within the inner region of SEI, indicating that this path is more energetically competitive. Overall, our enhanced-sampling MD simulations capture the formation of Li clusters within the SEI inner region, with distances about nano-meters scale away from the Li/SEI interface.

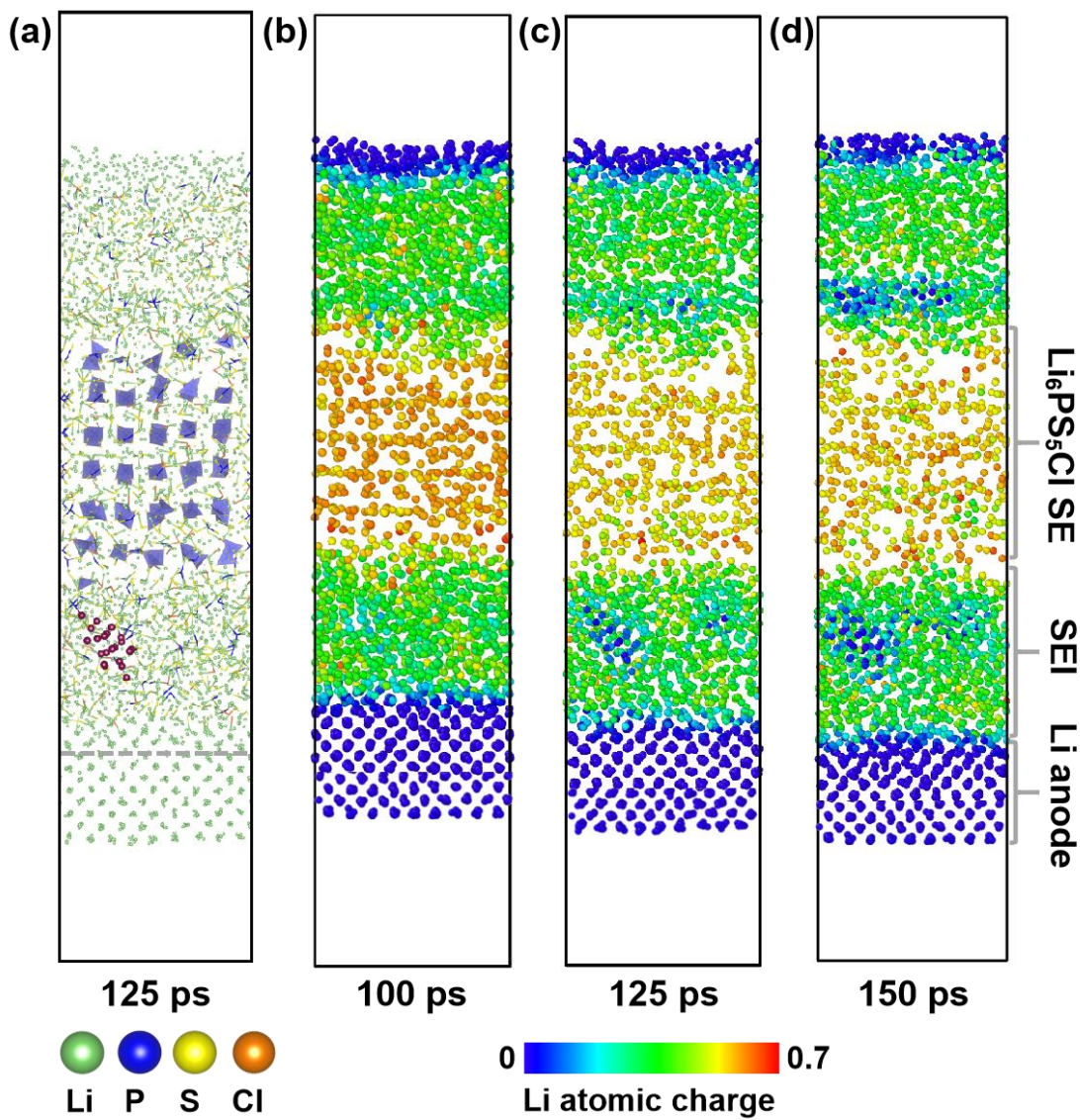

**Figure 3.** (a) Snapshot from enhanced-sampling DPMD simulations, at t = 125 ps. The emerged Li cluster driven by our enhanced-sampling approach is highlighted and colored in wine. The upper 3 layers of Li anode slab above the gray dashed line remain within the region of species B in the *CN* defined in Equation (1). (b-d) Li atomic charge distribution at 100, 125, and 150 ps. Only Li atoms are displayed and color-coded in panels (b), (c) and (d) for clarity of structural presentation. The Li atomic charges are calculated from the charge equilibration QEq method.

At the interface in a realistic battery, a Li cluster's formation within the SEI region implies that Li$^+$ ions need to be reduced by electrons (e$^-$) in SEI. As we acknowledge in the above Methods section that, we assume the artificial bias introduced in our enhanced-sampling MD approach could represent the electrochemical driving force for reducing Li ions. A question then naturally arises: could electrons transport from the Li metal anode to the Li cluster emergence region through the thin layer of SEI? If the electronic conductance is plausible, then the assumption of utilizing our designed $CV$ to represent electrochemical driving force could be rationalized. To resolve this concern, we calculate the system's bandgap, a fundamental parameter related to material's electronic conductivity. Our bandgap of a bulk-phase $Li_6PS_5Cl$ computed by the DFT method (with the PBE exchange-correlation functional) shows a value of 2.15 eV, in good agreement with previous work.[39-41] The bulk $Li_6PS_5Cl$ phase thus exhibits a wide gap and an electronic insulating feature,[42,43] which is unlikely to provide electronic conductive channels. Since the electrons need to transport from the Li metal anode to the inner side of SEI to reduce $Li^+$, the nanoscale SEI region between the Li anode surface and the Li clustering spot must show certain level of electronic conductivity. To verify this hypothesis, we further calculate the bandgap of the SEI.

As mentioned earlier, the SEI is a composite material, mainly composed of $Li_2S$, $Li_3P$, and LiCl. Therefore, we treat the SEI as an amorphous composite and extract its structure (outlined by the dashed rectangle in Figure 4a, with a side view shown in Figure 4c) from the 125 ps configuration in the enhanced-sampling DPMD simulations, when the Li cluster starts to emerge. We first perform structural optimizations for the extracted SEI layer using our trained DP force field (no vacuum included in this supercell and periodic along all three directions), and then calculate the bandgap of the SEI thin layer using the DFT approach (see SI for details). We find that the amorphous SEI thin layer exhibits zero bandgap, demonstrating its potentially high electronic conductivity. This is further illustrated by the density of states of the bulk $Li_6PS_5Cl$ and the extracted amorphous SEI layer, where the latter shows a noticeable bandgap decrease (Figure 4d). In addition, the thickness of our extracted SEI layer is only ~1 nm (see the rectangular region in Figure 4a), within the accessible range of electron tunneling.[44] Given the above evidences and analysis, the significantly reduced bandgap of the thin SEI enables promising electronic transport, and could reduce $Li^+$ ions to metallic Li atoms during the cluster formation process, which justifies the phenomenon of Li nucleation observed in our enhanced-sampling DPMD simulations.

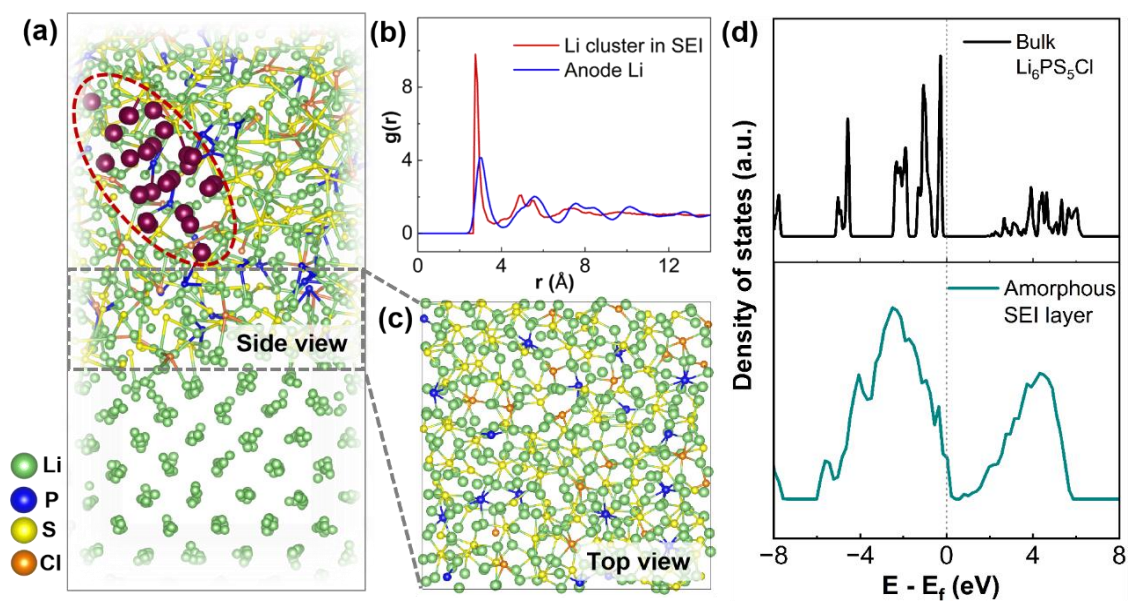

**Figure 4.** (a) Zoom-in atomic configuration of the snapshot at 125 ps with the emergence of a Li cluster (atoms highlighted in the red circle). We extract a slab from the SEI region (highlighted by the dashed rectangular). (b) The Li-Li radial distribution function (RDF) for the Li atoms in the emerged Li cluster (red) and anode Li (blue), respectively. (c) Top view of the extracted SEI slab. (d) Density of states of the bulk $Li_6PS_5Cl$ (top panel) and the extracted amorphous SEI layer (bottom panel) based on DFT calculations.

At this point, we have learned that the substantially reduced bandgap of the nanoscale amorphous SEI layer between the Li anode surface and Li clustering spots facilitates electron transport, enabling electron transport through the thin SEI layer and reducing $Li^+$ into Li cluster within the SEI. As these Li clusters grow, they may further develop into larger Li particles or even dendrites that penetrate through the SEI and the bulk of SE. Our results align with a recent experimental study by Yang and co-workers,[17] where they observed isolated dead Li clusters in the interior of $Li_6PS_5Cl$, and proposed that it could be caused by the cracking of SE.[10,17] We note that in a recent report from Gu et al.,[45] by treating $Li_2S$ crystal as SEI, the authors highlighted that the electron transport within the SEI layer is blocked due to the wide bandgap of $Li_2S$ (3.53 eV). Here in our report, we consider the SEI as a multi-component amorphous material, and show its zero-bandgap feature by DFT electronic structure calculations, indicating the importance of the SEI components and phases,[38,46,47] and further implying that tuning SEI compositions could significantly impact its electronic properties. Our results reveal the atomic-scale mechanism of the electronically disconnected Li clusters form away from the Li anode surface in a Li|LPSC|Li cell, which rationalizes recent experimental findings.[17] We therefore propose that in contrary to reducing the electronic conductivity of the bulk of SE, tailoring the electronic structure of the SEI is perhaps equally, if not more, important to prevent the initiation of lithium dendrites at the Li anode/SEI interface. To achieve, the chemistry of the SE and therefore its side reactions with the Li metal and formation chemistry of the SEI should be considered in the future. Furthermore, introducing structurally intact and rather thin artificial electron-insulating SEI such as LiF or other innovative compounds could be effective strategies to block the electron transport through the SEI and hence suppress extensive Li dendrite formation. More efforts could be dedicated to exploring these optimization approaches in the future.

## 4. Conclusion

To summarize, we have investigated the mechanism of the early stage of Li dendrite formation at Li|$Li_6PS_5Cl$ interfaces in ASSLBs. With an accurately trained machine-learning force field employed in enhanced-sampling DPMD simulations, a combination of coordination numbers as *CV* enables us to efficiently drive the kinetically sluggish process of Li cluster nucleation. We provide atomic-level insights into the dynamic paths of Li cluster nucleation and formation in ASSLBs. We show that Li clusters form within the SEI inner region, located ~1 nm away from the Li/SEI interface, rather than being directly connected to the Li-anode surface or within the SE bulk. A significantly reduced bandgap of an SEI interstitial region between the clustering spot and the anode surface facilitates electronic conduction, which could enable reduction of Li ions within the amorphous SEI. We reveal that the SEI near the Li anode/SE interface is an amorphous composite material, with its electronic properties determined by the SEI components and phases. Engineering the electronic properties of SEI components

(increasing its bandgap) thus is key to suppress Li dendrite formation in ASSLBs. For example, optimizing SEI compositions and phases to achieve an electronically insulating property while maintaining high ionic conductivity will be an ideal goal to inhibit Li dendrite growth, which could be explored by employing machine-learning assisted molecular simulations in future.

**Supporting Information**

Supporting Information is available from the Wiley Online Library or from the author. The file contains first-principles calculation details, construction of Deep Potential, atomic model used in enhanced-sampling DPMD simulations, details of enhanced-sampling DPMD simulations, and the charge equilibration QEq model.

**Acknowledgements**

The authors acknowledge funding support from the National Key Research and Development Program (grant no. 2021YFB2500200). The authors gratefully acknowledge funding support from the National Natural Science Foundation of China (grant no. 52273223, 92372115), DP Technology Corporation (grant no. 2021110016001141), the School of Materials Science and Engineering at Peking University, and the AI for Science Institute, Beijing (AISI). Y. A. acknowledges the financial support from the Postdoctoral International Exchange Program (no. YJ20220030) and the funding support from China Postdoctoral Science Foundation (no. 2023M730043). The Bohrium Cloud Platform supported by DP Technology is acknowledged for computational resources.

# Supporting Information

**Observing Li Nucleation at Li Metal–Solid Electrolyte Interface in All-Solid-State Batteries**

*Yun An,[1] Taiping Hu,[1,2] Quanquan Pang,[1*] and Shenzhen Xu[1,2*]*

[1]Beijing Key Laboratory for Theory and Technology of Advanced Battery Materials, School of Materials Science and Engineering, Peking University, Beijing 100871, China

[2]AI for Science Institute, Beijing 100084, China

Corresponding authors:
*qqpang@pku.edu.cn
*xushenzhen@pku.edu.cn

**Contents**

# 1. First-principles calculation details

First-principles calculations were carried out using the Vienna Ab initio Simulation Package (VASP)[1,2] based on density functional theory (DFT) with plane wave basis. Core-valence electrons were treated by the projector augmented wave (PAW) approach[3,4]. The semi-local generalized gradient approximation (GGA) Perdew-Burke-Ernzerhof (PBE)[5] was employed. The kinetic cutoff energy was set to be 600 eV. The energy and force convergence criteria were set as $10^{-6}$ eV and 0.01 eV/Å for the electronic and ionic steps in relaxation, respectively.

We first optimized the bulk structure of BCC-Li and $Li_6PS_5Cl$. Gamma-centered scheme of 11×11×11 and 2×2×1 k-point grids were employed for bulk BCC-Li and $Li_6PS_5Cl$, respectively. The optimized lattice constants were 3.42 Å for BCC-Li and 10.25 Å for $Li_6PS_5Cl$, in good agreement with previous reports.[6] These optimized structures were used in the initial dataset generation to train the deep potential.

The electronic structure of the extracted amorphous SEI layer (Figure 4c in the main text) was calculated in the bulk phase, using a Gamma-centered scheme of 1×1×3 k-point grids with 600 eV kinetic energy cutoff. The other setup and convergence criteria are the same as the above mentioned first-principles calculations.

# 2. Construction of Deep Potential

**2.1 Initial dataset**

Three different systems were used to construct the initial dataset: bulk BCC-Li, bulk $Li_6PS_5Cl$, and the $Li/Li_6PS_5Cl$ slab interface, containing 16, 52, and 185 atoms, respectively. After full structural relaxation, random perturbations were applied, followed by 20 ps ab initio molecular dynamics (AIMD) simulations with the NVT ensemble[7] at 300 K to generate initial data. The Nose-Hoover thermostat[8,9] was employed throughout the AIMD simulations during this stage.

**2.2 DP-GEN iterations**

Next, we employed the Deep Potential GENerator (DP-GEN)[10] concurrent learning scheme for model training. The DP-GEN scheme contains a series of successive iterations, each of which contains three stages: exploration, labeling, and training.[10]

During the exploration stage, the model deviations, defined as the maximal standard deviation of the atomic force predicted by the model, were estimated. By setting lower and upper boundaries for the trust levels of $\sigma_{low}$ and $\sigma_{high}$, when the model deviation $\epsilon$ of a structure falls within the range of [$\sigma_{low}, \sigma_{high}$], the structure will be considered a candidate and will be added to the dataset for the next iteration. DP-GEN is considered as converged when the predicted accuracy of the DP model was higher than 99%.[11] Using the trained DP model, DPMD simulations lasting 10-20 ps were performed with NVT ensemble, over a temperature range of 200 to 1000 K. We notice that the trust level bounds should be adjusted based on the variation in temperature and system. For example, in the low temperature range of 200 to 500 K, the values of [$\sigma_{low}, \sigma_{high}$] for bulk Li, bulk $Li_6PS_5Cl$, and the Li|$Li_6PS_5Cl$ interfacial system were set to [0.10, 0.25]. In the high temperature of 600 to 1000 K, the [$\sigma_{low}, \sigma_{high}$] for bulk Li and bulk $Li_6PS_5Cl$ were adjusted to [0.15, 0.30], while for the Li|$Li_6PS_5Cl$ interfacial system were set to [0.25, 0.50].

During the training stage, four models with random seeds were trained via the deepmd-kit package.[12] Both the embedding and fitting neuron networks consisted of three layers, with (25, 50, 100) and (240, 240, 240) nodes, respectively. The Adam stochastic optimization method[13] was applied to minimize the loss function, with an exponentially decaying learning rate from $1\times10^{-3}$ to $5\times10^{-8}$.

## 3. Atomic model used in enhanced-sampling DPMD simulations

To investigate the interfacial evolution of Li/$Li_6PS_5Cl$ with a large-size model in enhanced-sampling DPMD simulations, we employed the 3×3 (001) surface of

Li$_6$PS$_5$Cl and the 9×9 (001) surface of BCC-Li. This resulted in the supercell lengths of 30.814 Å for Li$_6$PS$_5$Cl and 30.763 Å for BCC-Li, indicating their negligible lattice mismatch (mismatch by only 0.16%). For the choice of the crystallographic directions in the slab model, as Li$_6$PS$_5$Cl was predicted to be a 3D ionic conductor,[14] we used the [001] crystallographic direction for both Li$_6$PS$_5$Cl and BCC-Li to construct the Li/Li$_6$PS$_5$Cl slab, in accordance with previous reports.[15,16]

We then constructed a large symmetric sandwich model of Li|Li$_6$PS$_5$Cl|Li, containing 4833 atoms in total. The bottom region in the model of Li|Li$_6$PS$_5$Cl|Li is a BCC-Li slab as an anode (16 Li layers); the middle region is the Li$_6$PS$_5$Cl electrolyte; and the upper material of BCC Li (9 Li layers) is included in our interfacial supercell to represent a source of Li particles from an assumed deeper SE region. A 40 Å of vacuum region was inserted between the anode Li slab and the Li source slab to mimic a half-cell interfacial system (see Figure 1c in the main text). The resulting Li|Li$_6$PS$_5$Cl|Li simulation model has dimensions of 30.81×30.81×150.42 Å$^3$. This system was then used in the enhanced-sampling DPMD simulations.

## 4. Details of enhanced-sampling DPMD simulations

We used the above constructed Li|Li$_6$PS$_5$Cl|Li system in the enhanced-sampling DPMD simulations via the interface of the deepmd-kit[12] within the LAMMPS package.[17] We employed the NPT ensemble condition in our simulations, allowing the volume to relax in response to an ambient pressure. We set the pressures to be 1 bar along the x and y directions while the cell's length along the z direction is fixed due to the presence of vacuum region (no pressure was set along this z direction). The periodic boundary condition was applied for all of the three directions. The temperature was set to 300 K. The timestep of 1 fs was employed to generate MD trajectories by time integration.

To expedite the emergence of Li cluster/dendrite formation in MD simulations, we employed a collective variable (*CV*)-based enhanced sampling strategy – the moving restraint method – as implemented in the PLUMED package.[18] Within this framework, a harmonic potential of $\frac{1}{2}\kappa(s(\vec{R}) - s_0(t))^2$, on defined *CV*, was used to drive the system toward a specific state. Here, $\kappa$ is the force constant, set as 500 eV; $s$ is the defined *CV* as a function of the system's atomic configuration $\vec{R}$, and the time-dependent $s_0(t)$ performs as a knob to move the *CV* along a specific direction at a specific rate.

We used atoms' coordination numbers (*CN*) to construct the *CV* in our enhanced-sampling DPMD simulations. The coordination numbers of Li-Li, Li-S, Li-P, and Li-Cl were used in the modeled system. The *CN* quantity[18] follows the switching function of

$$CN_{AB} = \sum_{i \in A} \sum_{j \in B} \frac{\left(1 - \frac{r_{ij}}{R_0}\right)^n}{\left(1 - \frac{r_{ij}}{R_0}\right)^m} \tag{S1}$$

where the $r_{ij}$ is the distance between the $i^{th}$ atom among the species A and $j^{th}$ atom among the species B; the exponents *n* and *m* determine the sharpness of the switching function; $R_0$ represents a spherical radius for defining a neighboring region of a centered atom (belonging to species A) when counting coordinated atoms (within species B)[18]. The detailed values of *n*, *m*, and $R_0$ are given in Table S1. The reference for $R_0$ in $CN_{Li-Li}$ was chosen based on the Li-Li distances between BCC-Li and $Li_6PS_5Cl$, while the references for $R_0$ in the $CN_{Li-S}$, $CN_{Li-P}$, and $CN_{Li-Cl}$ were selected from the Li-S, Li-P, and Li-Cl distances in $Li_6PS_5Cl$ (with $R_0$ values slightly larger than these distances).

**Table S1**. Parameters used to calculate coordination numbers (*CN*) of Li-Li, Li-S, Li-P, and Li-Cl pairs.

| $CN_{AB}$ | Parameters |
|---|---|

$$CN_{Li-Li} = \sum_{i \in Li} \sum_{j \in Li} \frac{(1 - \frac{r_{LiLi}}{R_0})^n}{(1 - \frac{r_{LiLi}}{R_0})^m} \quad r_{LiLi}: \text{distance between Li-Li atoms; } R_0 = 3.2 \text{ Å}; n = 6; m = 12$$

$$CN_{Li-S} = \sum_{i \in Li} \sum_{j \in S} \frac{(1 - \frac{r_{LiS}}{R_0})^n}{(1 - \frac{r_{LiS}}{R_0})^m} \quad r_{LiS}: \text{distance between Li-S atoms; } R_0 = 3.0 \text{ Å}; n = 6; m = 12$$

$$CN_{Li-P} = \sum_{i \in Li} \sum_{j \in P} \frac{(1 - \frac{r_{LiP}}{R_0})^n}{(1 - \frac{r_{LiP}}{R_0})^m} \quad r_{LiP}: \text{distance between Li-P atoms; } R_0 = 4.0 \text{ Å}; n = 6; m = 12$$

$$CN_{Li-Cl} = \sum_{i \in Li} \sum_{j \in Cl} \frac{(1 - \frac{r_{LiCl}}{R_0})^n}{(1 - \frac{r_{LiCl}}{R_0})^m} \quad r_{LiCl}: \text{distance between Li-Cl atoms; } R_0 = 4.0 \text{ Å}; n = 6; m = 12$$

## 5. The charge equilibration QEq model

In the QEq approach,[19] for a system containing $N$ atoms, the atomic charges $Q_i$ are optimized to minimize the QEq energy, $E_{\text{QEq}}$:

$$E_{\text{QEq}} = E_{\text{Coulomb}} + \sum_i^N (\chi_i^0 Q_i + \frac{1}{2} J_i Q_i^2) \tag{S2}$$

where $E_{\text{Coulomb}}$ is Coulomb interactions; $\chi_i^0$, $Q_i$, and $J_i$ are the electronegativity, atomic charge, and atomic hardness, respectively.

Since our system is periodic, we employed the Ewald summation algorithm[20] to calculate the $E_{\text{Coulomb}}$:

$$E_{\text{Ewald}} = E_{\text{real}} + E_{\text{recip}} + E_{\text{self}} \tag{S3}$$

For the real space part, we have:

$$E_{\text{real}} = \frac{1}{2}\sum_{i=1}^{N}\sum_{j,j\neq i}^{N_{\text{neigh}}} Q_i Q_j \frac{\text{erfc}(\frac{r_{ij}}{\sqrt{2}\eta})}{r_{ij}} \qquad (S4)$$

here, the sum taken over all neighboring atoms within the real-space cutoff radius $r_{\text{cut}}$. $r_{ij}$ represents the distance between atom $i$ and atom $j$, and $\eta$ is the width of the auxiliary charges. For a given $r_{\text{cut}}$ and error tolerance $\delta$, $\eta$ is defined as $\sqrt{-log2\delta}/r_{\text{cut}}$, consistent with the OpenMM toolkit.[21]

For the reciprocal space part:

$$E_{\text{recip}} = \frac{2\pi}{V}\sum_{\mathbf{k}\neq 0} \frac{\exp(-\frac{\eta^2|\mathbf{k}|^2}{2})}{|\mathbf{k}|^2} \left|\sum_{i=1}^{N} Q_i \exp(i\mathbf{k}\cdot\mathbf{r}_i)\right|^2 \qquad (S5)$$

here, $V$ is the volume of the unit cell. The sum is taken over all reciprocal lattice points within the reciprocal space cutoff radius.

For the self-interaction part:

$$E_{\text{self}} = -\sum_{i=1}^{N} \frac{Q_i^2}{\sqrt{2\pi}\eta} \qquad (S6)$$

We employed the Particle-Mesh Ewald (PME)[20] algorithm to calculate Coulomb interactions. The PME energy, $E_{\text{PME}} \equiv E_{\text{real}} + E_{\text{recip}} + E_{\text{self}}$, is calculated via the DMFF package.[22]

Given that the Gaussian charge distributions are used in the QEq calculations, a Gaussian charge correction should be applied, following the expression:

$$E_{\text{corr}}^{\text{Gauss}} = -\frac{1}{2}\sum_{i=1}^{N}\sum_{j\neq i}^{N_{\text{neigh}}} Q_i Q_j \frac{\text{erfc}\left(\frac{\sqrt{\sigma_i^2+\sigma_j^2}}{\sqrt{2}\gamma_{ij}}\right)}{\sqrt{\sigma_i^2+\sigma_j^2}} + \sum_{i=1}^{N} \frac{Q_i^2}{2\sqrt{\pi}\sigma_i} \qquad (S7)$$

here, $\sigma_i$ represents the width of Gaussian charge density, determined by the relevant elements' covalent radii.

Due to the atomic charges may induce a significant dipole, particularly in the direction perpendicular to the interface, therefore, a dipole correction[23] term should be included:

$$E_{\text{corr}}^{\text{dipole}} = \frac{2\pi}{V}\left(\left(\sum_i^N q_i z_i\right)^2 - \sum_i^N q_i \sum_i^N Q_i z_i^2 - \left(\sum_i^N q_i\right)^2 \frac{L_z^2}{12}\right) \quad (S8)$$

here $L_z$ represents the box length along the direction perpendicular to the Li/Li$_6$PS$_5$Cl interface, and $z_i$ represents the z component of the $i^{\text{th}}$ atom's coordinate.

As shown in Equation (4) in the main text, the total atomic charges in our system were constrained to 0, requiring constrained minimization. This was achieved using the Lagrange multiplier method, as discussed in the main text.

Given all the terms discussed above, the QEq energy of the system is described by the following form:

$$E_{\text{QEq}} = \sum_i^N (\chi_i^0 Q_i + \frac{1}{2} J_i Q_i^2) + E_{\text{PME}} + E_{\text{corr}}^{\text{Gauss}} + E_{\text{corr}}^{\text{dipole}} \quad (S9)$$

The electronegativity and hardness parameters used in the QEq calculations are taken from earlier references[24,25] as shown in Table S2.

**Table S2.** Electronegativity and hardness parameters used in our QEq calculations.

|  | Li | P | S | Cl |
|---|---|---|---|---|
| Electronegativity | -3.0000 | 1.8000 | 6.5745 | 10.0000 |
| Hardness | 10.0241 | 7.0946 | 9.0000 | 6.0403 |

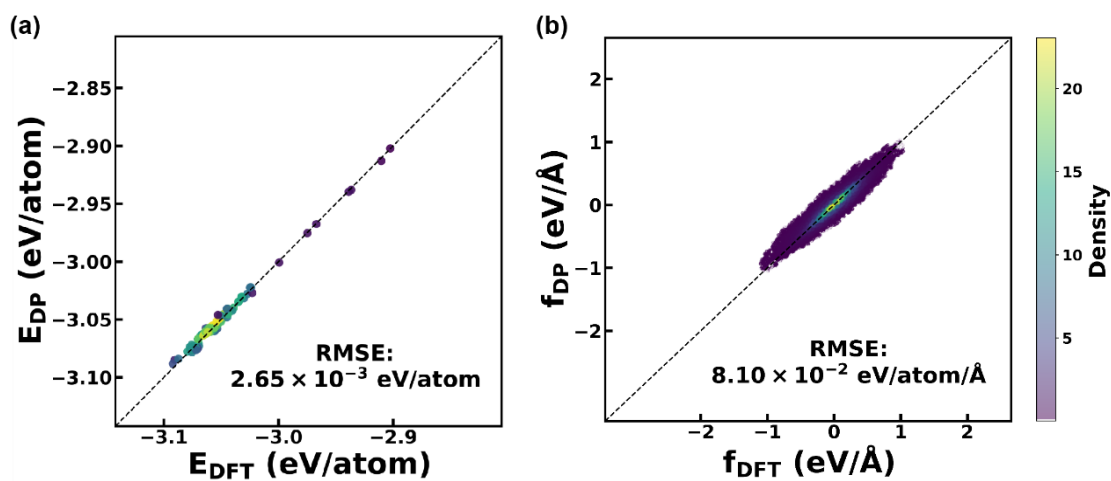

**Figure S1**. Energies (a) and forces (b) predicted from the DP force field model and DFT on the training dataset.